# The social aspects of quantum entanglement

***Confessions of a theoretical physicist.*** Our readers might be wondering how and why an article with such a strange title fits in a journal of popular culture. Actually, the common reaction when people meet a theoretical physicist is to look at him like an alien, or an autistic psychopath unable to conduct a normal life enjoying more 'popular' subjects. Well, there are many reasons why theoretical physics is not so popular even among people with a passion for culture. Everybody would be puzzled when hearing the absurdities that we often profess about how Nature works! Quantum theory, specifically, affirms that our everyday experience of the "macroscopic" world (things that we see and touch) is useless to describe how the "microscopic" world (the minuscule components of the same things that we see and touch) behaves. These tiny objects don't even know whether they are particles (like billiard balls) or waves of energy (like sun rays), and not because we are unable to detect them with enough precision, but because there is an ultimate *a priori* limit to the knowledge that can be gained. Thus, depending on how we look at them, they behave as particles or as waves, answering randomly to our tests.

Now go and tell everybody that the best minds of the last century discovered this (confirmed by over 70 years of experimental evidences): that the world is intrinsically random and it's impossible to have complete information of its components, and this implies that there are for example poor kittens which are at the same time dead and alive… I guess they will feel a bit depressed, and/or will think that you are completely drunk!! Indeed, quantum theory might come handy! This "uncertainty" principle says that if you know perfectly where a particle is, you cannot know its speed, and viceversa. So next time a policeman stops you saying you're running too fast with your car, try to tell him (and let us know if it works) that according to a Mr Heisenberg, as you are in a given place there's no way to know which was your speed; and if he insists that he has recorded your speed, you answer that you couldn't be in that given place at that moment, as you are delocalized in the entire universe!!!

Now, if a couple of brave readers arrived alive at this point, they must have at least a terrible headache! Relax, you are in good company. Mr Einstein eventually refused quantum ideas because "God – he said – does not play dice". And Mr Feynman (and I'm mentioning all Nobel Prize winners!) honestly said: "Nobody understands quantum mechanics". So, do I have any hope to make quantum physics a popular piece of culture? I will now tell you something about the most counterintuitive and radical feature of quantum mechanics.. yet I will try to make it as natural as the feeling that everybody experiences from the moment of birth and before: love.

***Entanglement: passion at a distance.***[1] Mr Schrödinger (another pioneer of quantum mechanics, and Nobel Prize winner: he promoted the cat example) coined the term "entanglement" (entrelazamiento) in 1935 to describe a peculiar connection between quantum systems:

> *"When two systems enter into temporary physical interaction due to known forces between them, and when after a time of mutual influence the systems separate again, then they can no longer be described in the same way as before. I would not call that one but rather <u>the</u> characteristic trait of quantum mechanics, the one that enforces its entire departure from classical lines of thought. By the interaction the two systems have become **entangled**."*

Entanglement thus manifests as a somehow puzzling correlation (Einstein blamed it as a "spooky action at a distance") between parties who once came into contact, and mantain their contact even miles away. This has been experimentally demonstrated with individual atoms or light beams: but how can it fit in our everyday experience of life? The closest feeling which comes into my mind is

---

[1] This expression was coined by R. D. Gill.

*love*. Think of a mother and a child, or two lovers who shared an intense emotion, and are now living at the opposite sides of the world. They *feel* each other, perceive the happiness or the sadness of the distant partner, and are influenced by this.

Schrödinger added:

> *"Another way of expressing the peculiar situation is: the best possible knowledge of a whole does not necessarily include the best possible knowledge of all its parts. The lack of knowledge is due to the interaction itself."*

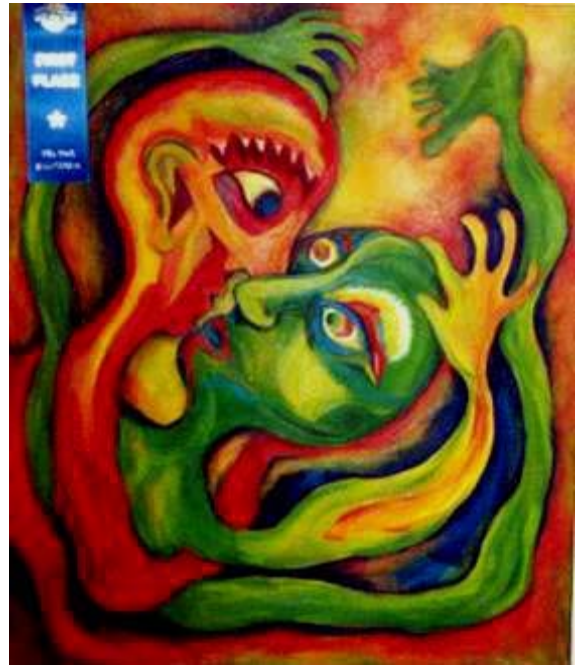

*"Entanglement"* (58x71cm, acrylic)
© Pamela Ott, 2002 <**www.hottr6.com/ott**>
*reproduced with permission*

In our metaphor, nobody of the two lovers is complete on its own. Only when taken together, they complement each other. *They are non-separable halves of the same entangled entity.* No proper and complete understanding, on both physical and psychological grounds, is available for this phenomenon. But the language of art, probably, can make it clearer: entanglement is admirably depicted by Pamela Ott, who has almost zero knowledge of quantum mechanics (I asked her!) and paints "from her subconscious". The waveness of the lines, the choice of complimentary colors, the faded entwining of bodies and souls is what in my opinion most closely resembles a true image of entanglement, and of loving passion.

***Entangled families: monogamy versus promiscuity.*** Now the subject becomes intriguing. What happens if there are three or more parties in this love game? Technically, we are addressing the complex topic of "multipartite entanglement". In this context, there's a fundamental law: entanglement is *monogamous*! If our two lovers, traditionally called Alice and Bob, really love each other a lot – they are maximally entangled – there's no way for Alice to be even a little bit in love with a Carlos (or the same for Bob and a Clara). This is maybe a bit controversial, as in our modern culture it's becoming more and more popular to violate monogamy in human relationships…but at least let me say that entanglement, now that we look at it with the eyes of love, is not that bad beast that people (even eminent ones) thought it to be. Are you starting to like it? Be advised though that if the love between Alice and Bob is not total, each of them can share some love with extra parties!! So back home, be sure that your partner is *maximally* entangled to you, otherwise, suspicions are legitimate!!

Another interesting aspect can arise in multi-party entangled relationships. The above picture (total entanglement excludes third parties) strictly applies to very simple, 'binary' objects (or persons), only capable of answering either yes or no to any question without more structured argumentations (we call them "qubits"). But suppose now that Alice and Bob are more complex characters (we call them "continuous-variable systems"), capable to answer questions providing a series of elaborate motivations. Such kind of more reliable persons, may at some point think to enrich their family life with a baby. Can they love their baby while being monogamous to each other? *Yes*, they can! In this scenario, the entanglement (love) between any two of them in their family can be arbitrarily strong, and this enhances the genuine entanglement (love) shared by the three of them (mom, dad, and the baby). Isn't it a nice romantic picture? Actually, an alternative description could be in terms of a group of three or more people, which might also be all of the same sex! And this "freedom" that arises if their personality is multifaceted and not binary, is a sort of *promiscuity*: the more any two of them are attracted by each other, the more they like to stay altogether acting in threesomes and more intricate orgies! The quantum society, you see, is not that much different from ours after all: we are made of quanta and immersed into them… who's influencing who?

***Is there a serious quantum impact on our society?*** These are just funny ways to try to relate the microscopic world (in which entanglement has probably nothing to do with love) to the world in which we are used to live. Everybody may imagine the situation which in his fantasy best applies to this strange quantum realm. Already at the ages of Romans and Greeks, people invented figures (gods) associated to natural events, and pictured them as banqueting and being involved in love affairs. Entanglement is surely one of the mysterious 'gods' of the Book of Nature as we know it today, and "when it comes to entanglement – according to V. Vedral – we've only just discovered the tip of an iceberg".

I conclude by hinting at what entanglement can really do for us. Apart from its quite interesting social aspects at the microscopic level, the new technological *quantum information* revolution is taking place, based on entanglement and its properties. In particular, Alice and Bob can send flirting emails to each other with the unconditional security that nobody can eavesdrop, as they encrypt their data via a secret key established from the intrinsic randomness of quantum particles, and protected by the monogamy of entanglement. And this is reality: quantum cryptosystems are plug-and-play devices which you can buy from MagiQ <[www.magiqtech.com/](www.magiqtech.com/)> or IdQuantique <[www.idquantique.com](www.idquantique.com)>, and install on your PC to definitely solve the problem of (quantum) hackers. Similarly, we are working to build quantum computers, which exploit quantum superpositions (wave and particle, dead and alive..) to do parallel calculations billions of times faster than current supercomputers!!! And there is much, much more…

I've enjoyed speaking about one particular aspect of science, which is entering into and gradually affecting popular culture. Like this, there are many others. Now, next time you meet a theoretical physicist, please don't look at him in that strange way :) and remember … we are all entangled!

*Gerardo Adesso*

*Università degli Studi di Salerno (Italy)*
*& Universitat Autònoma de Barcelona (Spain)*

[gerardo.adesso@gmail.com](gerardo.adesso@gmail.com)